\documentclass[authoryear,preprint,review,12pt]{elsarticle}

\usepackage{amssymb}

\usepackage{lineno}

\journal{Earth and Planetary Science Letters}

\begin{document}

\begin{frontmatter}



\title{Size of the group IVA iron meteorite core: Constraints from the age and composition of Muonionalusta}


\author[label1]{Nicholas A. Moskovitz\corref{cor1}}

\address[label1]{Carnegie Institution of Washington, Department of Terrestrial Magnetism, 5241 Broad Branch Road, Washington, DC 20015, USA}

\author[label2]{Richard J. Walker}

\address[label2]{University of Maryland, Department of Geology, College Park, MD 20742}

\cortext[cor1]{Corresponding author: nmoskovitz@dtm.ciw.edu}

\begin{abstract}
The group IVA fractionally crystallized iron meteorites display a diverse range of metallographic cooling rates. These have been attributed to their formation in a metallic core, approximately 150 km in radius, that cooled to crystallization in the absence of any appreciable insulating mantle. Here we build upon this formation model by incorporating several new constraints. These include (i) a recent U-Pb radiometric closure age of $<$2.5 Myr after solar system formation for the group IVA iron Muonionalusta, (ii) new measurements and modeling of highly siderophile element compositions for a suite of IVAs, and (iii) consideration of the thermal effects of heating by the decay of the short-lived radionuclide $^{60}$Fe. Our model for the thermal evolution of the IVA core suggests that it was approximately 50 - 110 km in radius after being collisionally exposed. This range is due to uncertainties in the initial abundance of live $^{60}$Fe incorporated into the IVA core. Our models define a relationship between cooling rate and closure age, which is used to make several predictions that can be tested with future measurements. In general, our results show that diverse cooling rates and early U-Pb closure ages can only coexist on mantle-free bodies and that energy released by the decay of $^{60}$Fe reduces the core size necessary to produce diverse metallographic cooling rates. The influence of $^{60}$Fe on cooling rates has largely been neglected in previous core formation models; accounting for this heat source can affect size estimates for other iron meteorite cores that cooled to crystallization in the presence of live $^{60}$Fe. Candidates for such a scenario of early, mantle-free formation include the iron IIAB, IIIAB and IVB groups.
\end{abstract}


\end{frontmatter}

\linenumbers

\section{Introduction}
\label{intro}

Amongst the oldest melted rocks in the solar system, fractionally crystallized iron meteorites provide insight on the early stages of planet formation. These meteorites are interpreted as fragments of cores from planetesimals that melted and subsequently differentiated due to heating by the decay of short-lived radioactive isotopes like $^{26}$Al and $^{60}$Fe \citep{Mittlefehldt98,Goldstein09,Moskovitz11}. Though these meteorites are highly evolved, differences in their composition and structure provide a basis for classification \citep{Haack05}. Fourteen well-defined groups of iron meteorites have been identified, most of which are thought to represent the cores of distinct parent bodies \citep{Goldstein09}. While the classification of these groups is generally agreed upon, details of their formation, such as parent body size, are less certain.

The origin of the group IVA iron meteorites has long been debated due to several unusual properties \citep{Willis78,Rasmussen95,Haack96,Scott96,Wasson01,Ruzicka06,Wasson06,Yang08}. First, they display the widest range of metallographic cooling rates (100-6600 K/Myr) of any iron meteorite group \citep{Yang08}. These rates were recorded during the formation of the Widmanst\"{a}tten pattern (WP, a structure of interleaved bands of kamacite and taenite) as the core cooled from 1000 to 700 K. In addition, the diameters of cloudy zone (CZ) particles at the boundaries of taenite crystals suggest that the IVA cooling rates between 600 to 500 K varied by a factor of fifteen \citep{Yang07}. Second, several IVAs contain silicate inclusions, which have been explained through a host of impact and melt evolution scenarios \citep{Ulff95,Haack96,Ruzicka06,Wasson06,McCoy11}. Third, they are significantly depleted in moderately volatile siderophile elements relative to chondrites and other iron groups \citep{Wasson01,Yang08,McCoy11}. Lastly, though not exclusive to the IVAs, their trace element abundances are consistent with sampling a majority of their parent core's fractional crystallization sequence \citep{Scott96,Wasson01,Ruzicka06,McCoy11}. 

One formation scenario that reasonably explains these properties (see \citet{Yang08} and \citet{Ruzicka06} for overviews of other models) suggests that the differentiated IVA parent body was originally $\sim$1000 km in diameter prior to a hit-and-run collision with a comparably massive proto-planet \citep{Asphaug06,Yang07}. This collision released a string of metal-rich fragments that eventually cooled to crystallization without the insulating effects of an overlying silicate mantle. In this scenario, the isolated core that would eventually be disrupted to produce the IVA meteorites must have been $\sim$300-km in diameter to reproduce the wide range of metallographic cooling rates. A lack of insulating mantle is necessary to achieve rapid cooling ($>$1000 K/Myr) near the surface and the large size ensures that the center of the body cools slowly ($\sim$100 K/Myr). 

This formation scenario merits revisiting in light of a recent U-Pb age for the IVA Muonionalusta, which indicates system closure (i.e. cooling below $\sim600$ K) at 4565.3 $\pm$ 0.1 Ma \citep{Blichert10}. This is the earliest measured age for any fractionally crystallized iron. It falls only $\sim1$ Myr after the basaltic angrite Asuka 881394, the oldest differentiated rock in the Solar System \citep{Wadhwa09}, and is less than 2.5 Myr after the formation of CAIs (calcium-aluminum-rich inclusions), generally considered to be the first solids to condense from the solar nebula. The aim of this paper is build upon the IVA formation model first presented in \citet{Yang07} by taking into account newly available constraints on the evolution of the IVA parent core. In \S\ref{geochem} we present a fractional crystallization model to provide context for the origin of Muonionalusta relative to other IVAs. In \S\ref{thermal} we outline a thermal conduction model that includes heating by the decay of $^{60}$Fe. This thermal model and the result of the fractional crystallization calculation are used to constrain the size of the IVA core and make predictions for the closure ages of other IVAs (\S\ref{results}). In \S\ref{discussion} we discuss the sensitivity of our results to various assumptions inherent to the thermal model. In \S\ref{conclusions} we summarize and discuss the broader implications of this work.

\section{Model of IVA Fractional Crystallization}
\label{geochem}

\begin{figure}[]
\begin{center}
\includegraphics[width=14cm]{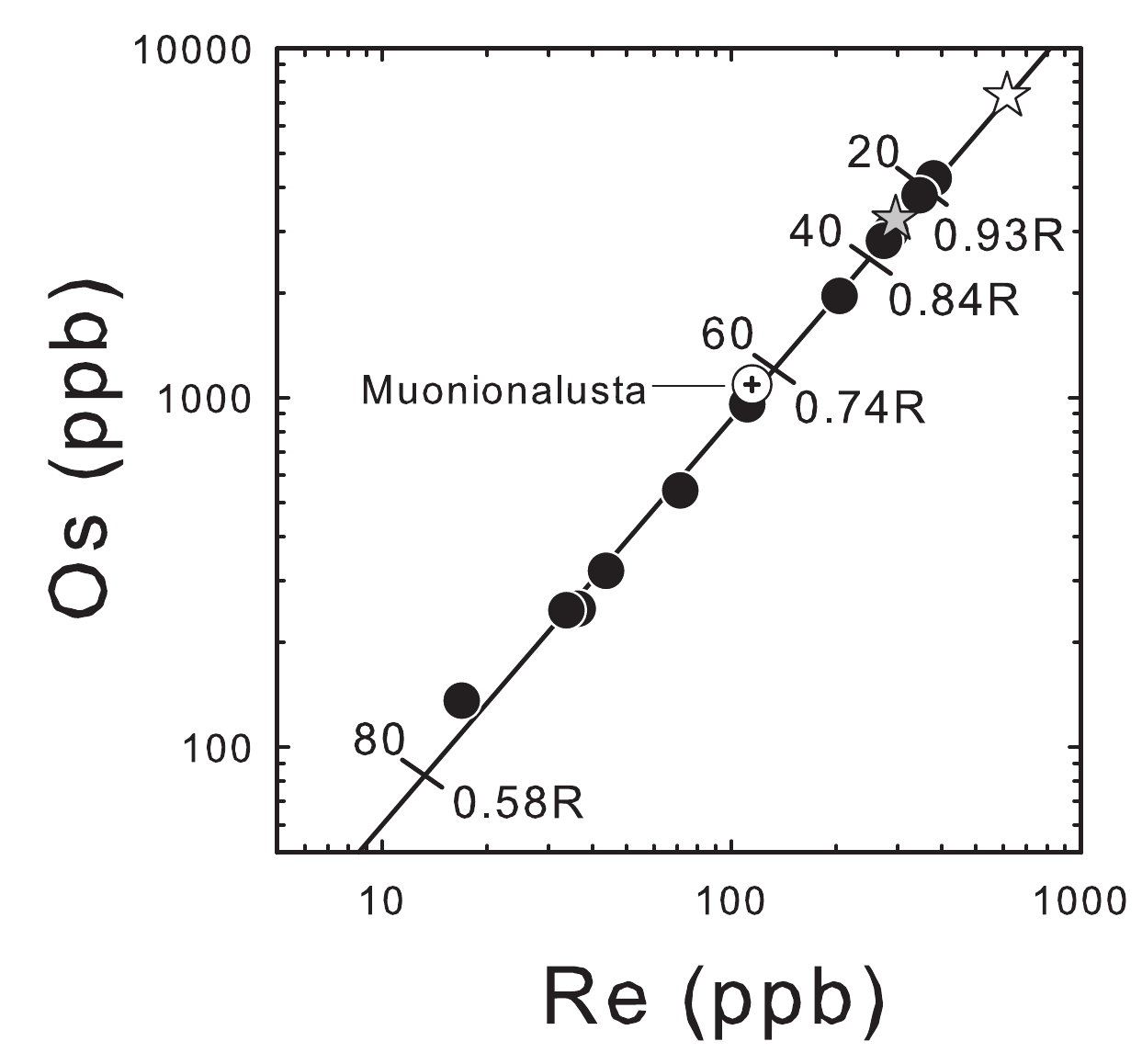}
\end{center}
\caption[]{Plot of Re versus Os (in ppb) for 14 group IVA iron meteorites. Muonionalusta is shown by the open circle with a cross. The solid line is the fractional crystallization trend for 50:50 mixes of equilibrium solids and liquids, using parameters discussed in the text. Tick marks indicate 20 through 80\% extents of fractional crystallization (equivalent to 0.58-0.93R in an inwardly crystallizing metallic core). For this model, Muonionalusta is produced after $\sim$60\% fractional crystallization. The grey star represents the assumed initial liquid composition; the open star denotes the composition of the first solid to form from this liquid.} 
\label{fig.muon}
\end{figure}

To estimate the extent of fractional crystallization required to produce Muonionalusta we model the IVA system assuming 3\% and 0.1\% initial S and P, respectively, using an approach similar to that of \citet{Walker08} for the IVB irons (Fig. \ref{fig.muon}). Initial Re and Os concentrations of 295 and 3250 ppb, respectively, are estimated for the IVA parental melt. Details of this model are provided in \citet[][]{McCoy11}. For these starting parameters, the Re and Os concentrations of Muonionalusta are attained after $\sim$60\% fractional crystallization, assuming that it has a composition consistent with an equilibrium solid. Appropriate concentrations are attained after 50\% fractional crystallization if Muonionalusta has a composition consistent with that of the evolving IVA liquid. Mixtures of solid and liquid compositions are achieved by intermediate extents of fractional crystallization. This result is generally consistent with the model of \citet{Wasson01}, who generated a composition similar to Muonionalusta after $\sim$40\% and $\sim$55\% fractional crystallization, using Au versus Ir and As versus Ir plots, respectively. It is also consistent with \citet{Yang08} who showed that an iron with 8.4\% Ni-content like Muonionalusta would crystalize after $\sim$60\% solidification of a core with initial 3 wt.\% S.

For an inwardly crystallizing core, as expected for the IVA parent \citep{Haack92,Ruzicka06,Yang08}, these data suggest that Muonionalusta formed between approximately 80-70\% of the parent body radius, with 70\% representing the best fit to our data (Fig. \ref{fig.muon}). This sub-surface origin is consistent with Muonionalusta's Ni abundance (an indicator of cooling rate), which is intermediate to other IVAs \citep{Blichert10,Yang08}.

\section{Thermal Conduction Model}
\label{thermal}

Because of the need for rapid cooling, our model, like that of \citet{Yang07}, begins with a body without any insulating silicate mantle. This model is based on the 1D thermal conduction equation \citep{Moskovitz11} and assumes a metallic sphere with the following properties: density 7500 kg/m$^3$, thermal conductivity 50 W/m/K, specific heat 400 J/kg/K, initial uniform temperature of 1750 K, and a fixed boundary temperature of 200 K. These material properties are within 20\% of those used in other thermal models for differentiated bodies and are generally applicable to iron alloys \citep{Ghosh98,Ghosh99,Hevey06,Yang07,Moskovitz11}. The initial and boundary temperatures match those used by \citet{Yang07}. We use a fixed-temperature, Dirichlet boundary condition, which is simpler to implement than a radiative boundary condition, does not introduce significant error for the range of temperatures in which we are interested, and produces results no different from a radiative boundary for all radii up to a few km from the surface \citep{Ghosh98,Yang07,Moskovitz11}. The boundary at 200 K is an approximation for the ambient temperature in the solar nebula \citep{Yang07}. In \S\ref{temps} we show that our results are insensitive to a reasonable range of assumed initial and boundary temperatures.
 
One difference between our model and those of previous investigators is that we have considered heating by the decay of $^{60}$Fe. Previous models \citep[e.g.][]{Haack90,Yang08} assumed that no heat sources were available after $^{26}$Al caused melting and differentiation. However, the solar system's initial abundance of $^{60}$Fe relative to its stable isotope $^{56}$Fe is not precisely known. Therefore we present two scenarios for the IVA parent core: one in which the core thermally evolves with a maximum possible abundance of $^{60}$Fe and the second in which no $^{60}$Fe is present. These scenarios produce lower and upper limits respectively to the size of the IVA core. After accounting for a recently revised half-life of 2.62 Myr and properly reduced mass spectrometry results, a reasonable maximum for the ratio $^{60}$Fe/$^{56}$Fe at the time of CAI formation is $4\times10^{-7}$ \citep{Rugel09,Mishra10,Telus11,Ogliore11}. For the maximum heating case, we adopt this half-life and abundance, along with a decay energy of 3.04 MeV and an Fe mass fraction of 90\% for the IVA parent body \citep{Ghosh98,Mittlefehldt98}. This case assumes that all of the chondritic $^{60}$Fe is sequestered into the core, again resulting in an upper limit on the energy available from radiogenic decay.

Other studies suggest that the initial $^{60}$Fe/$^{56}$Fe ratio was up to several orders of magnitude lower and/or heterogeneously distributed in the solar nebula, resulting in reservoirs largely depleted in $^{60}$Fe \citep{Chen09,Quitte10,Spivak11,Tang11}. To bracket the range of possible abundances we also consider the thermal evolution of the IVA parent core in the absence of any $^{60}$Fe. This minimum-$^{60}$Fe model is analogous to that presented by \citet{Yang08}, though we now use this model to match the additional constraints of age and formation radius for Muonionalusta. The specific lower limit for $^{60}$Fe is not critically important; once $^{60}$Fe/$^{56}$Fe drops more than an order of magnitude below the upper limit, i.e. $\leq5\times10^{-8}$, the energy available from the decay of $^{60}$Fe would have been insignificant in the overall thermal budget of the IVA parent body.

The thermal model assumes that the IVA core was fully formed at 4567.7, a weighted mean of recently measured CAI ages \citep{Jacobsen08,Burkhardt08}. The statistical uncertainty on this weighted mean is 0.3 Myr, though systematic errors could be as large as 1 Myr \citep[see \S\ref{precision}, ][]{Amelin10}. This assumption requires that accretion, differentiation and exposing of the IVA core occurred much faster than the $\sim$10$^6$-year timescales for $^{60}$Fe heating and conductive cooling \citep{Rugel09,Moskovitz11}. This is consistent with hydrodynamic simulations of planetesimal formation in turbulent proto-planetary disks that show bodies as large as 1000 km accreted in much less than 10$^3$ years \citep{Johansen07}. In \S\ref{delay} we show that this nearly instantaneous accretion anytime within 1.5 Myr of CAI formation, a limit for fractionally crystallized iron cores, does not affect our size estimates for the IVA core \citep{Haack05,Qin08}. The assumed initial temperature of 1750 K at 4567.7 Ma is unphysical, however a variety of heat sources (e.g. $^{26}$Al decay and impacts) could have contributed to the global heating of bodies hundreds of km in size \citep{Keil97,Moskovitz11}. These heat sources would have been relevant on timescales of 10$^5-10^6$ years.

\section{Models of the IVA Core}
\label{results}

\begin{figure}[]
\begin{center}
\includegraphics[width=9cm]{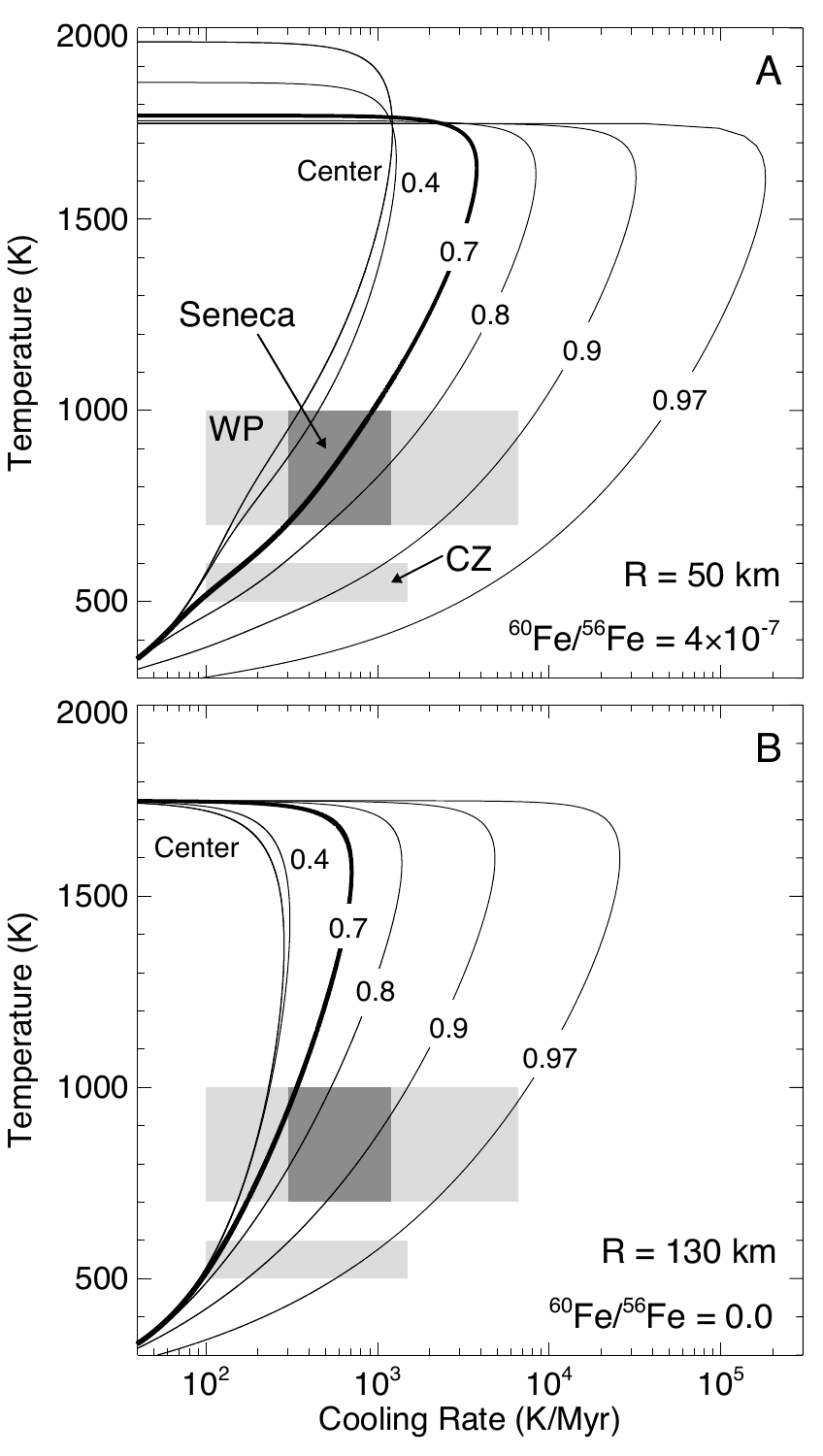}
\end{center}
\caption[]{Cooling rate versus temperature at different depths, expressed as fractions of the parent radius, in exposed cores with maximal $^{60}$Fe (A) and in the absence of $^{60}$Fe (B). The light grey regions represent temperatures and measured IVA cooling rates during formation of the Widmanst\"{a}tten pattern (WP) and cloudy zone particles (CZ). The dark grey box represents the measured cooling rates for Seneca Township, a sample with similar Ni-content to Muonionalusta. In each case the sizes of the cores have been adjusted such that the measured ranges of WP and CZ rates are reproduced, and so that a radius of 0.7R crystallizes at the rates measured for Seneca. These two examples represent upper (R=130 km) and lower (R=50 km) limits to the size of the IVA core.} 
\label{fig.cr}
\end{figure}

Figure \ref{fig.cr} shows how temperature varies as a function of cooling rate in two different exposed cores. These temperature profiles are shown for a range of depths, from the center of the body up to 97\% of the radius $R$. The top panel (A) depicts a case with maximum $^{60}$Fe/$^{56}$Fe=$4\times10^{-7}$ and $R=50$ km; the bottom panel (B) depicts a core with no $^{60}$Fe and $R=130$ km.

The release of energy from the decay of $^{60}$Fe prolongs cooling and reduces cooling rates. Thus, the body in Figure \ref{fig.cr}A, which is only 50 km in radius, exhibits cooling rates as small as 140 K/Myr at its center during the formation of the Widmanst\"{a}tten pattern (1000-700 K) and a range of rates during the formation of cloudy zone particles  (600-500 K) that is consistent with measurements. Furthermore, WP cooling rates at 0.7R are as expected for Muonionalusta based on rates measured for its compositional analog Seneca Township \citep[300-1200 K/Myr,][]{Yang08}. If no $^{60}$Fe were present on a body this small then none of these constraints would be met: the lowest cooling rates during WP formation would be an order of magnitude larger, the CZ rates would only vary by a factor of 10, and the WP cooling rates at 0.7R would be larger than those measured for Seneca Township.  Smaller bodies with this  $^{60}$Fe abundance cool too quickly to reproduce the low end of the IVA range of WP rates. As such, a radius of 50 km is a lower limit for a core to be able to reproduce the IVA cooling rate data.

An upper limit for the size of the IVA core is depicted in Figure \ref{fig.cr}B. In this case, no $^{60}$Fe is present, thus requiring $R=130$ km to achieve slow cooling ($\sim100$ K/Myr) near the center. A body with these properties meets the WP, CZ and Seneca cooling rate constraints. Although a larger body would produce the necessary range of WP and CZ rates, the cooling rates at 0.7R would be too slow ($<100$ K/Myr) for Seneca Township. Therefore, the cooling rates modeled in Figure \ref{fig.cr} bracket the size of the IVA core to somewhere between 50 and 130 km.

The age of Muonionalusta further constrains the size of the IVA core. Figure \ref{fig.closure} shows the times at which different depths in exposed cores reach U-Pb isotopic closure at 600 K \citep{Blichert10}. The two panels again correspond to $^{60}$Fe/$^{56}$Fe=$4\times10^{-7}$ (A) and $^{60}$Fe/$^{56}$Fe=0 (B) and produce lower and upper size limits respectively. For the former, Muonionalusta would crystallize at 4565.3 Myr in a core with a radius of 55 km (i.e. where the bold curve intersects the shaded region). For the later, the age of Muonionalusta is reproduced for a core 110 km in radius.

Matching both the cooling rates (Fig. \ref{fig.cr}) and the U-Pb age at the expected depth of Muonionalusta's origin (Fig. \ref{fig.closure}) suggests that the IVA core was between 50 and 110 km in radius, depending upon the assumed initial abundance of $^{60}$Fe. Below this lower limit the entire body would reach U-Pb closure within 2.5 Myr (Fig. \ref{fig.closure}A), requiring that Muonionalusta come from the center of the IVA core. This would unreasonably preclude cooling rates slower than the 500 K/Myr inferred for Muonionalusta \citep{Blichert10}. Conversely, a core with R$>$110 km would not reach U-Pb closure in $<$2.5 Myr at 0.7R (Fig. \ref{fig.closure}B). We adopt R=50 km as a lower limit rather than 55 km because several of the assumptions in these models hasten cooling and thus demand an increase in size to produce slow $\sim100$ K/My cooling (see \S\ref{discussion}).

In these models, size and $^{60}$Fe abundance are degenerate properties whose variation produces a series of solutions that fulfill the cooling rate and age constraints. The examples in Figures \ref{fig.cr} and \ref{fig.closure} are end-member cases that bracket the range of possibilities. If the initial $^{60}$Fe abundance in the IVA parent core were known, then it is possible to predict a specific relationship between U-Pb closure age and cooling rate. For example, a given depth profile in Figure \ref{fig.cr} records a range of cooling rates between 1000-700 K and a U-Pb closure age at 600 K. Table \ref{tab.ages} presents predicted ages for several IVAs by assuming that the mean cooling rate measured for each sample corresponds to the rate midway between 1000 and 700 K. These predictions are specific to the case of R=50 km and $^{60}$Fe/$^{56}$Fe=$4\times10^{-7}$ and are simply intended to highlight several implications of this formation model.

First, the results in Table \ref{tab.ages} predict a wide range of U-Pb closure ages from 4564.0 - 4567.5 Ma. This range of several Myr is a general outcome of these models, irrespective of the assumed $^{60}$Fe abundance, and is resolvable with the current precision of U-Pb dating techniques \citep{Blichert10}. None of our models produce rates less than $\sim$140 K/Myr and thus cannot predict the closure ages of the four most chemically evolved IVAs (Duchesne, Chinautla, New Westville and Steinbach), though it is likely they reached U-Pb closure after 4654 Ma. We do not view this as a significant failure of the model considering the large uncertainties and many assumptions that have been made (see \S\ref{discussion}). Varying input parameters such at the specific heat could reduce the calculated cooling rates to the 100 K/My that is measured for Duchesne.

Second, the correlation between closure age and cooling history predicts cooling rates for Muonionalusta. For example, when R=50 km and $^{60}$Fe/$^{56}$Fe=$4\times10^{-7}$, Muonionalusta's closure age suggests cooling rates between 290-940 K/Myr, which by design are similar to the range of rates for Seneca Township \citep{Yang08}. 

Lastly, this formation model predicts that IVAs with the fastest cooling rates will have the oldest U-Pb closure ages. In the example shown in Table \ref{tab.ages}, the predicted ages for samples like La Grange and Bishop Canyon are only a few hundred thousand years separated from the formation of CAIs. If this formation scenario is correct, then these IVAs  may represent the oldest differentiated rocks in the Solar System.

\section{Assumptions in the Thermal Model}
\label{discussion}

\subsection{Neglected processes}
\label{neglect}
The release of latent heat during crystallization, the temperature dependence of specific heat, and the effect of an insulating layer have been neglected in these models. The net latent heat available in the Fe-Ni-S system is 270 kJ/kg \citep[][]{Haack90}, which is less than the total energy released by the decay of $^{60}$Fe when the initial $^{60}$Fe/$^{56}$Fe$>6\times10^{-8}$. Unfortunately this critical $^{60}$Fe abundance is intermediate in the range of current best estimates \citep{Chen09,Mishra10,Quitte10,Telus11,Spivak11,Tang11}. As such, it is difficult to determine the relative importance of latent heat release without a well-constrained $^{60}$Fe abundance. Nevertheless, a release of 270 kJ/kg of latent heat could increase temperatures by up to several hundred Kelvin (assuming no conductive loss of heat and a heat capacity of 400 J/kg/K) and, like any additional heat source, would prolong cooling. For all models presented here, the temperature-independent specific heat is near the lower limit for metallic iron and thus hastens cooling \citep{Ghosh99}. 

To quantify the effects of an insulating mantle we have run a series of simulations for a 50-km body with $^{60}$Fe/$^{56}$Fe=$4\times10^{-7}$ and a surface layer with low thermal diffusivity. These calculations suggest that a silicate mantle (thermal conductivity = 2 W/m/K, heat capacity = 1200 J/kg/K) thicker than $\sim$0.5 km would prevent rapid cooling $>$1000 K/Myr for nearly all radii. With an initially chondritic composition, the core should be of order 40\% of the radius, i.e. 20 km in this example, overlain by a mantle $\sim$30 km thick \citep{Haack90}. Therefore, any reasonable thickness for a silicate mantle would prevent the rapid cooling rates ($>$1000 K/Myr) measured for some IVAs. This argues strongly in favor of a collision exposing the IVA core before it cooled below 1000 K \citep{Yang07}.

We have performed a similar calculation for a surface regolith (thermal conductivity = 0.01 W/m/K, heat capacity = 1200 J/kg/K) in place of a mantle \citep{Haack90}. A regolith thicker than $\sim$50 m would reduce the range of cooling rates to below that measured for the IVAs. For the range of regolith and mantle thickness considered, 0-0.5 km and 0-5 km respectively, the dominant effect on the thermal evolution is a reduction of the highest cooling rates at the base of the insulating layer. These insulating layers do not influence central cooling rates until they become sufficiently thick (approximately 0.5 km and 5 km respectively) such that the timescale over which they conductively transfer heat becomes comparable to the timescale of conductive heat loss across the whole of the 50-km metallic core.

Neglecting the release of latent heat, the temperature dependence of specific heat, and the effect of an insulating layer, act to hasten the loss of thermal energy and thus artificially increase interior cooling rates. Since the slowest IVA cooling rates are most difficult to reproduce in a mantle-free body, treatment of these phenomena could extend the estimated range of possible parent bodies to sizes less than R=50 km.

Our model does not treat the convection of molten material, which would occur for melt fractions $>50\%$. For the Fe-Ni-S system, the assumed initial temperature of 1750 K would correspond to high degrees of partial melting. The details of convective processes in an exposed molten core are not well understood and are beyond the scope of this study. Convection in such a system would increase the rate of heat loss from the interior and perhaps require larger bodies to ensure $\sim100$ K/Myr cooling near the center.

\subsection{Initial and Boundary Temperatures}
\label{temps}
The ability of our model to reproduce the IVA cooling rates is unaffected by changes to initial temperatures down to 1000 K and nebular temperatures up to 300 K. For instance, the range of cooling rates during the formation of the Widmanst\"{a}tten pattern (1000-700 K) from the center to 0.9R for the R=50 km body in Figure \ref{fig.cr}A is 140-8500 K/Myr. Reducing the initial temperature of this body from 1750 K to 1000 K changes this range to 140-16000 K/Myr, only the fastest cooling rates near the surface are affected. The reason for the increase in cooling rates at the upper end of this range is due to our focus on the narrow range of temperatures at which the Widmanst\"{a}tten pattern forms. In other words, the lowering of initial temperature shifts the curves in Figure \ref{fig.cr} downwards, which causes a wider range of rates between 1000-700 K. Increasing the boundary temperature up to a plausible maximum of 300 K \citep{Yang07} also has little effect on cooling rates and results in a range of 110-6500 K/Myr for the above example. The thermal model of \citet{Yang08} is similarly insensitive to changes in initial and boundary temperatures.

\subsection{U-Pb Closure Temperature}
\label{closure}
These models assume a U-Pb closure temperature of 600 K, however this value is not well constrained. \citet{Blichert10} suggest 573 K based on the assumption that the closure temperature of Pb in sulfides is the same as that of Os. Studies of U and Pb diffusion in silicates suggest closure temperatures of about 625$\pm50$ K \citep{Spear96}. This suggests that our adopted closure temperature of 600 K may be accurate to approximately $\pm100$ K, but specific measurements of U/Pb diffusion in sulfides at elevated temperatures are essential for interpreting the U-Pb ages of iron meteorites. This uncertainty predominantly affects the calculated closure ages in Figure \ref{fig.closure}. For example, a lower closure temperature of 500 K demands a slightly smaller body (approximately 50 km rather than 55 km when $^{60}$Fe/$^{56}$Fe$=4\times10^{-7}$) to match the age constraint for Muonionalusta. This difference is insignificant relative to other uncertainties in the model.

\subsection{Precision of U-Pb Ages}
\label{precision}
We have assumed that the closure age of Muonionalusta is 4565.3 $\pm$ 0.1 Ma \citep{Blichert10}. However, this value could be in error by $\sim1$ Myr due to variations in the $^{238}$U/$^{235}$U value of CAIs \citep{Amelin10}. Ages younger by 1 Myr would permit larger parent bodies because of the additional time available to reach U-Pb closure. Older ages would require smaller bodies such that closure occurs even faster. Therefore, this uncertainty translates into a slightly broader range of acceptable parent body sizes. At the lower size limit this effect is insignificant, i.e.~less than 5 km difference from our standard assumption of CAI ages equal to 4567.7 Ma. At the upper limit, a younger closure age for Muonionalusta could allow parent bodies up to R=125 km.

\subsection{Effect of Delayed Accretion}
\label{delay}
Hf-W chronometry and thermal modeling suggest that fractionally crystallized iron meteorite parent bodies accreted within 1.5 Myr of CAI formation \citep{Qin08}. We adopt this constraint as an upper limit to the time of formation and use it to recalculate the results in Figures \ref{fig.cr} and \ref{fig.closure}. For Figure \ref{fig.cr}A, the range of cooling rates for radii inside of 0.9R are 140-8500 K/Myr. Delaying accretion to 1.5 Myr has little effect on these rates, it simply reduces the amount of live $^{60}$Fe and produces a range of cooling rates from 225-9000 K/Myr. Figure \ref{fig.cr}B is unaffected by the assumed time of accretion due to a lack of $^{60}$Fe. In short, later times of accretion primarily act to dampen peak temperatures and have little influence on cooling rates below 1000 K.

The size constraints for the IVA core are similarly unaffected by times of accretion up to 1.5 Myr. Delayed accretion effectively shifts the curves in Figure \ref{fig.closure} upwards, producing no change in the upper limit for the core size. In other words, the point of intersection of the 0.7R curve and the age of Muonionalusta will never shift to larger radii with later times of accretion. Delayed accretion also has little effect on the lower size limit. This insensitivity is a result of competing processes in cores heated by $^{60}$Fe. With delayed accretion less time is available to reach U-Pb closure, thus requiring faster cooling to produce Muonionalusta at 0.7R. But, less live $^{60}$Fe is present to prolong cooling. These two effects act to cancel out one another.

\section{Conclusions}
\label{conclusions}

We have argued that the parent core of the iron IVA meteorites was between 50 - 110 km in radius, dependent primarily on the abundance of live $^{60}$Fe incorporated into the core. Cores in this size range follow a crystallization sequence that match constraints set by metallographic cooling rates \citep{Yang07,Yang08}, a U-Pb age \citep{Blichert10} and geochemical data \citep{Rasmussen95,Wasson01,Ruzicka06,McCoy11}. An estimate of R=150$\pm50$ km by \citet{Yang08} is consistent with the upper end of this size range, though our models are also consistent with a IVA core up to three times smaller. The lower gravity in a smaller parent may facilitate the trapping of silicate inclusions in a molten core \citep{Ruzicka06}. The probability of near-catastrophic collisions necessary to expose a molten core should be greater for smaller bodies \citep{Bottke05}. The dependence of our results on a single old age emphasizes the need for additional dating of Muonionalusta and other IVAs. Experimental confirmation of Muonionalusta's predicted cooling rates of $\sim500$ K/Myr is equally important to future studies on the thermal evolution of the IVA core.

The range of cooling rates for the IVAs is the greatest of any chemical group \citep{Haack05,Yang06,Goldstein09,Yang10}, suggesting they represent one of the largest iron meteorite parent cores with measured rates. Determination of cooling rates for currently unmeasured groups could reveal otherwise. With a radius between 50 - 110  km, the IVA core must have derived from a fully differentiated parent body that was at least twice as large \citep{Haack90}. However, due to large uncertainties regarding the hit-and-run formation scenario, it is difficult to precisely estimate the size of the fully differentiated IVA parent body prior to the collision. Nevertheless, our results suggest that planetary bodies $\sim$200-500 km in diameter were present during the first few Myr of solar system history, with a possibility of even larger bodies depending upon the details of the collision that exposed the IVA core and the currently unknown cooling rates of other iron groups. This removes the necessity, though does not preclude the possibility of $10^3$ km, proto-planetary bodies early in solar system history \citep{Yang08}. 

For this scenario of formation within an exposed core,  IVAs with fast cooling rates ($>$1000 K/Myr) may have absolute ages separated by as little as a few times 10$^5$ years from CAI formation (Table 1). Confirmation of this would provide important clues to understanding the timescales involved in the formation of the first planetary bodies in the solar system.

\begin{figure}[]
\begin{center}
\includegraphics[width=9cm]{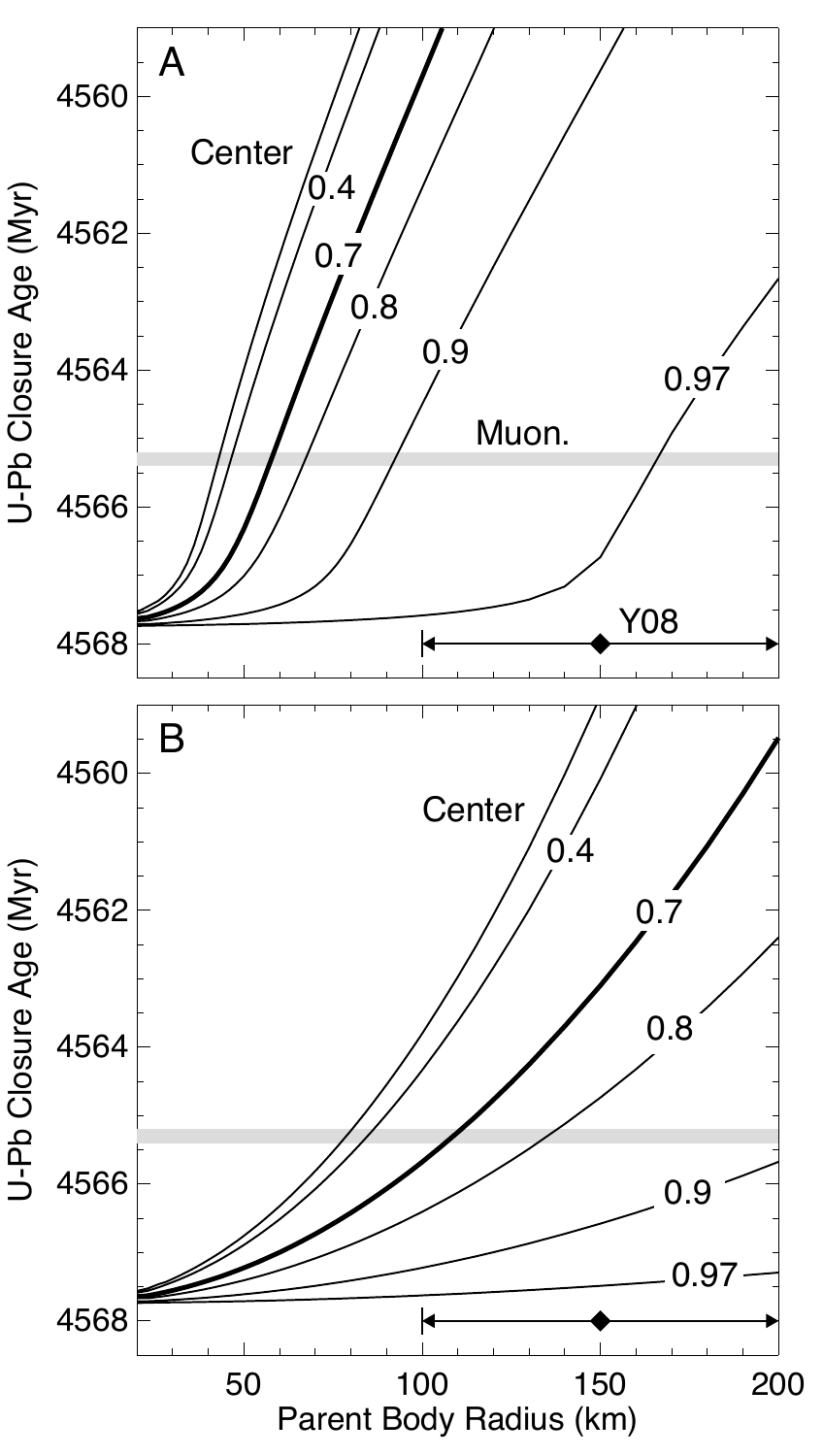}
\end{center}
\caption[]{U-Pb closure ages at different depths for a range of parent body sizes with initial $^{60}$Fe/$^{56}$Fe=$4\times10^{-7}$ (panel A) and $^{60}$Fe/$^{56}$Fe=0 (panel B). The thick curve (0.7R) represents the expected radius of Muonionalusta's origin. The grey region denotes Muonionalusta's closure age of 4565.3$\pm$0.1 Ma. The size estimate for the IVA parent body from \citet{Yang08} is shown at the bottom right. The closure age at the formation radius of Muonionalusta is reproduced for core radii between 50 (panel A) and 110 km (panel B), depending upon the initial abundance of $^{60}$Fe.
} 
\label{fig.closure}
\end{figure}

Though Muonionalusta experienced shock melting, one of its troilite grains somehow preserved the old age measured by \citet{Blichert10}. If similar troilite grains can be dated in other IVAs, then the resulting range of U-Pb closure ages will provide a tighter constraint on the size of the IVA parent. For example, the range of closure ages for an R=50-km core is less than half that of an R=100-km core (Fig. \ref{fig.closure}). The core sizes of other fractionally crystallized iron groups with diverse cooling rates, such as the IIAB, IIIAB and IVBs \citep{Yang08,Yang10}, can be constrained by modeling both cooling rates and U-Pb ages, assuming such ages can be measured.
 
Previous studies on the formation and crystallization of fractionally crystallized irons assumed that the thermal effects of $^{60}$Fe were insignificant. This is reasonable for irons that formed inside of cores surrounded by insulating mantles since they would crystallize after $^{60}$Fe was largely extinct. However, the age of Muonionalusta suggests that mantle-free cores may have crystallized when $^{60}$Fe was still extant. As we have shown, cooling rates can be affected by the decay of this isotope. Unfortunately, current uncertainties in its initial abundance make it difficult to specifically quantify the relevance of $^{60}$Fe decay to the thermal evolution of these parent bodies. If $^{60}$Fe was present during the crystallization of iron groups like the IVA, IIAB, IIIAB and IVBs, then updated models will result in a reduction of the inferred sizes of these parent bodies.\\

{\bf Acknowledgements} We thank Eric Gaidos and Rick Carlson for helpful comments regarding this study. Ed Scott, Joe Goldstein and an anonymous referee provided thoughtful reviews that led to significant improvement of this manuscript. NAM acknowledges support from the Carnegie Institution of Washington and the NASA Astrobiology Institute. This work was in part supported by NASA Cosmochemistry grant NNX10AG94G to RJW.

\begin{table}
\begin{center}
\begin{tabular}{lcc}
\hline 
\hline
& Cooling Rate$^a$ & Predicted U-Pb  \\
Meteorite &  (K/Myr) & Closure Age (Ma) \\
\hline
La Grange		&	6600		&	4567.5  \\
Obernkirchen		&	2900		&	4567.4  \\
Bishop Canyon		&	2500		&	4567.4  \\
Jamestown		&	1900		&	4567.3  \\
Gibeon			&	1500		&	4567.2  \\
Seneca Township	&	580		&	4566.4  \\
Altonah			&	420		&	4565.8  \\
Muonionalusta		&	290-940$^b$	&	4565.3$\pm$0.1$^c$ \\
Bushman Land		&	260		&	4564.5  \\
Duel Hill			&	220		&	4564.0  \\
Steinbach			&	150		&	$<4564$ \\
New Westville		&	120		&	$<4564 $ 	\\
Chinautla			&	110		&	$<4564 $  \\
Duchesne			&	100		&	$<4564 $ 	 \\
\hline
\end{tabular}
\caption[]{Cooling rates and U-Pb closures ages for a IVA core with R=50 km and $^{60}$Fe/$^{56}$Fe=$4\times10^{-7}$\\
$^a$ \citet{Yang08}\\
$^b$ Model prediction\\
$^c$ \citet{Blichert10}}
\label{tab.ages}
\end{center}
\end{table}

\end{document}